\documentclass[acmsmall,screen,nonacm]{acmart}

\usepackage{thmtools}
\usepackage{mathpartir}
\usepackage{listings}
\usepackage{amsmath}
\usepackage{mathtools}

\usepackage{macros}

\begin{document}

\title{Remarks on Algebraic Reconstruction of Types and Effects}

\author{Patrycja Balik}
\email{pbalik@cs.uni.wroc.pl}
\author{Szymon Jędras}
\email{sjedras@cs.uni.wroc.pl}
\author{Piotr Polesiuk}
\email{ppolesiuk@cs.uni.wroc.pl}
\affiliation{%
  \department{Institute of Computer Science}
  \institution{University of Wrocław}
  \city{Wrocław}
  \country{Poland}
}

\begin{abstract}
  In their 1991 paper ``Algebraic Reconstruction of Types and Effects,''
  Pierre Jouvelot and David Gifford presented a type-and-effect
  reconstruction algorithm based on an
  algebraic structure of effects.
  Their work is considered a milestone in the development of type-and-effect
  systems, and has inspired numerous subsequent works in the area of
  static analysis.
  However, unlike the later research it spawned, the original algorithm
  considered a language with higher-rank polymorphism, a feature which
  is challenging to implement correctly.
  In this note, we identify subtle bugs related to variable binding in their
  approach to this feature.
  We revisit their type system and reconstruction algorithm,
  and describe the discovered issues.
\end{abstract}

\maketitle

\section{Introduction}

In 1991, Pierre Jouvelot and David Gifford published the seminal paper
``Algebraic Reconstruction of Types and
Effects''~\cite{DBLP:conf/popl/JouvelotG91}.
Their work was the foundation of many later results in the area of static
analysis (\cite{%
  DBLP:journals/iandc/TalpinJ94,%
  DBLP:conf/popl/TofteT94,%
  DBLP:conf/ccl/NielsonN94,%
  DBLP:conf/popl/BirkedalTV96,%
  DBLP:conf/sas/FahndrichA97,%
  DBLP:journals/toplas/TofteB98,%
  DBLP:conf/fase/NielsonAN98,%
  DBLP:journals/toplas/HelsenT04,%
  DBLP:journals/pacmpl/Elsman24}, to name a few),
in which effect information was used internally by the compiler, but
not exposed directly to the programmer.
However, the original language and algorithm considered by Jouvelot and
Gifford treated its effect system as a full-fledged user-facing mechanism,
with explicit effect polymorphic lambdas and higher-rank effect polymorphic
functions.
They have thus anticipated a feature that gained broader interest
only much later, \emph{e.g.}, in the context of practical type-and-effect
systems for algebraic effects with handlers~\cite{%
  DBLP:journals/corr/PlotkinP13,%
  DBLP:journals/corr/Leijen14,%
  DBLP:conf/haskell/WuSH14,%
  DBLP:journals/jlp/BauerP15,%
  DBLP:conf/icfp/HillerstromL16,%
  DBLP:journals/pacmpl/BrachthauserSO20,%
  DBLP:journals/pacmpl/BiernackiPPS19,%
  DBLP:journals/pacmpl/BiernackiPPS20,%
  DBLP:journals/pacmpl/XieCIL22,%
  DBLP:conf/onward/Madsen22,%
  DBLP:conf/esop/VilhenaP23}.

The combination of higher-rank polymorphism and type (and effect)
reconstruction is well-known to be tricky on theoretical grounds.
It requires a certain amount of type annotations to remain within the realm of
decidability~\cite{%
  DBLP:conf/popl/OderskyL96,%
  DBLP:conf/icfp/BotlanR03,%
  DBLP:journals/jfp/JonesVWS07,%
  DBLP:conf/icfp/DunfieldK13,%
  DBLP:journals/pacmpl/DunfieldK19,%
  DBLP:conf/pldi/EmrichLSCC20}.
But even with that hurdle out of the way, designing an algorithm to work in
this setting is faced with challenges involving the correct handling of
the ubiquitous type-variable binders, which is notoriously error-prone.
Theorems establishing properties of any such algorithm help assuage worries
about its soundness and completeness relative to a declarative type system
serving as the specification.
However, a pen-and-paper proof may not be sufficient, because small
variable-scoping problems are easy to miss when verified by humans.
Mechanized proofs provide a greater degree of trust, but tools such as
proof assistants took considerable time to reach sufficient maturity for
widespread use within the programming languages community.
Even the classic type reconstruction algorithm W was first formalized in
1999~\cite{%
    DBLP:journals/jar/NaraschewskiN99,%
    DBLP:journals/jar/DuboisM99},
and required rather tedious management of variable binding.

It is therefore no surprise that the proofs within ``Algebraic Reconstruction
of Types and Effects'' had to be verified by hand at the time.
We have identified subtle variable-related bugs due to which the algorithm
rejects certain valid programs, and in some cases infers incorrect
types.\footnote{%
  The problems pertain to the higher-rank polymorphism and
  explicit polymorphism features present in the calculus. The validity of
  later results building on this algorithm remains unthreatened, because
  they omitted these mechanisms.}
In this note, we relate the original type system and the reconstruction
algorithm, and describe our findings regarding the aforementioned issues.

\section{The Type System}

In this section we briefly recall the syntax and the type system.
As the original presentation uses LISP-like notation,
we adapt it to a more modern convention.
Moreover, we slightly change the way of representing variable binders
to more hygienic, but still equivalent forms.
We elaborate on this in \autoref{sec:differences}.

\subsection{Syntax}

\begin{figure}[t]
  \begin{alignat*}{2}
    & \tvarA, \tvarB, \tvarC, \ldots \span\span
      \tag{type variables} \\
    & \unifVarX, \unifVarY, \unifVarZ, \ldots \span\span
      \tag{unification variables} \\
    & \ident
        &&\Coloneqq \tvarA
        \mid \unifVarX
      \tag{identifiers} \\
    & \varX, \varY, \varZ, \varF, \ldots \span\span
      \tag{term variables} \\
    & \kind
        &&\Coloneqq \keffect
        \mid \ktype
      \tag{kinds} \\
    & \effect
        &&\Coloneqq \tvarA
        \mid \unifVarX
        \mid \effpure
        \mid \effect \effjoin \effect
      \tag{effects} \\
    & \type
        &&\Coloneqq \tvarA
        \mid \unifVarX
        \mid \type \tarrow{\effect} \type
        \mid \tforall{\tvarA}{\kind} \type
      \tag{types} \\
    & \descriptor
        &&\Coloneqq \effect \mid \type
      \tag{descriptors} \\
    & \expr
        &&\Coloneqq \varX
        \mid \lam{\varX}{\type}\expr
        \mid \lamU{\varX}\expr
        \mid \expr\;\expr
        \mid \lamD{\tvarA}{\kind}\expr
        \mid \expr\appD{\descriptor}
      \tag{expressions}
  \end{alignat*}
  \caption{The syntax.}
  \label{fig:syntax}
\end{figure}

The syntax is presented in \autoref{fig:syntax}.
We distinguish between type variables and
unification variables, and introduce the syntactic
category of identifiers, which are either of the two.
Since identifiers can represent both types and effects,
we have two kinds to distinguish between them:
the kind of all types $\ktype$ and the kind of all effects $\keffect$.
Effects are built from type variables, unification variables,
the pure effect $\effpure$,
and the join of two effects $\effect_1 \effjoin \effect_2$.
Types consist of type variables, unification variables,
effect-annotated arrow types $\type_1 \tarrow{\effect} \type_2$,
and polymorphic types $\tforall{\tvarA}{\kind} \type$, which bind
a type variable of a given kind in the type body.
Descriptors are either effects or types.

The syntax of expressions includes term variables $\varX$,
lambda abstractions with type annotations $\lam{\varX}{\type}\expr$,
and without type annotations $\lamU{\varX}\expr$,
function applications $\expr_1\;\expr_2$,
type abstractions $\lamD{\tvarA}{\kind}\expr$,
and type applications $\expr\appD{\descriptor}$.
Intuitively, the argument type of an unannotated lambda abstraction
is to be inferred, but with the restriction that it must be \emph{monomorphic},
\emph{i.e.}, it cannot contain any polymorphic types.
Formally, monomorphic types are defined by the following inductive relation.
\begin{mathpar}
  \inferrule*[right=MT-Var]
    { }
    {\monotype{\tvarA}}

  \inferrule*[right=MT-UnifVar]
    { }
    {\monotype{\unifVarX}}

  \inferrule*[right=MT-Arrow]
    {\monotype{\type_1} \and \monotype{\type_2}}
    {\monotype{\type_1 \tarrow{\effect} \type_2}}
\end{mathpar}

\subsection{Differences from the Original Presentation}
\label{sec:differences}

The presented syntax slightly differs from the original,
but in aspects that in our view improve readability
and do not change the essence of the system.
Here we briefly describe these differences.

\paragraph{Variables.}

In the original presentation,
binders in types were allowed to bind identifiers (type variables or unification
variables) by the grammar.
Since binders of unification variables can never occur in source programs
or the output of the algorithm, we decided to disallow this in the grammar.
The original presentation also reuses the syntactic category of identifiers for
term variables.
Since term variables are never mixed with type or unification variables,
we also treat them as a separate syntactic category.

The original grammar of identifiers allows
for \emph{constants}. Constants were used in two ways:
as global variables defined outside of the program,
and as fresh identifiers generated by the algorithm to represent
unknown abstract types introduced by polymorphic quantifiers.
In our presentation we use regular type and term variables for both purposes,
thus avoiding the need for a separate syntactic category.

\paragraph{Pure effect.}

In the original presentation, there was one distinguished constant
representing the pure effect.
Since we treat constants as regular type variables, we decided to represent
the pure effect as a separate construct $\effpure$ in the syntax of effects.

\paragraph{Monotypes.}

Originally, monomorphic types were defined by a separate syntactic category,
which included identifiers and arrow types.
Since we are biased towards formalizations in proof assistants,
we would like to think of separate syntactic categories
as separate inductive types.
Thus, we defined monomorphic types as an unary relation on
(polymorphic) types.

\subsection{Type System}

\begin{figure}[t]
  \begin{mathpar}
    \inferrule*[right=K-Var]
      { (\kindExt{\tvarA}{\kind}) \in \env }
      { \kinding{\env}{\tvarA}{\kind} }

    \inferrule*[right=K-Arrow]
      { \kinding{\env}{\type_1}{\ktype}
        \and
        \kinding{\env}{\type_2}{\ktype}
        \and
        \kinding{\env}{\effect}{\keffect} }
      { \kinding{\env}{\type_1 \tarrow{\effect} \type_2}{\ktype} }

    \inferrule*[right=K-Forall]
      { \kinding{\env, \kindExt{\tvarA}{\kind}}{\type}{\ktype} }
      { \kinding{\env}{\tforall{\tvarA}{\kind} \type}{\ktype} }

    \inferrule*[right=K-Pure]
      { }
      { \kinding{\env}{\effpure}{\keffect} }

    \inferrule*[right=K-Join]
      { \kinding{\env}{\effect_1}{\keffect}
        \and
        \kinding{\env}{\effect_2}{\keffect} }
      { \kinding{\env}{\effect_1 \effjoin \effect_2}{\keffect} }
  \end{mathpar}
  \caption{Kinding rules.}
  \label{fig:kinding}
\end{figure}

We start the presentation of the type system with the kinding rules
presented in \autoref{fig:kinding}.
The role of the kinding relation is twofold.
First, it ensures that types and effects are well-formed,
\emph{i.e.}, all type and effect variables are bound in the environment.
Second, as the relation is defined on descriptors, it allows us to
distinguish between types and effects, when needed.

The next ingredients of the type system are
the type equivalence ($\typeEquiv{\type_1}{\type_2}$)
and effect equivalence ($\effEquiv{\effect_1}{\effect_2}$) relations.
The effect equivalence relation is defined as the least congruence
containing the laws of an idempotent commutative monoid,
with $\effpure$ as the identity element and $\effjoin$ as the binary operation.
On top of the effect equivalence relation,
we define the type equivalence relation as the least congruence
that respects the effect equivalence relation in arrow types,
and allows to $\alpha$-rename bound type and effect variables
in polymorphic types.
As these definitions are standard, we omit them here.

\begin{figure}[t]
  \begin{mathpar}
    \inferrule*[right=T-Var]
      { (\varX : \type) \in \env }
      { \typing{\env}{\varX}{\type}{\effpure} }

    \inferrule*[right=T-Lam]
      { \omitted{\kinding{\env}{\type_1}{\ktype}}
        \and
        \typing{\env, \varX : \type_1}{\expr}{\type_2}{\effect} }
      { \typing{\env}{\lam{\varX}{\type_1}\expr}
          {\type_1 \tarrow{\effect} \type_2}{\effpure} }

    \inferrule*[right=T-LamU]
      { \omitted{\kinding{\env}{\type_1}{\ktype}}
        \and
        \monotype{\type_1}
        \and
        \typing{\env, \varX : \type_1}{\expr}{\type_2}{\effect} }
      { \typing{\env}{\lamU{\varX}\expr}
          {\type_1 \tarrow{\effect} \type_2}{\effpure} }

    \inferrule*[right=T-App]
      { \typing{\env}{\expr_1}{\type_2 \tarrow{\effect} \type_1}{\effect_1}
        \and
        \typing{\env}{\expr_2}{\type_2}{\effect_2} }
      { \typing{\env}{\expr_1\;\expr_2}
          {\type_1}{\effect_1 \effjoin \effect_2 \effjoin \effect} }

    \inferrule*[right=T-LamD]
      { \typing{\env, \kindExt{\tvarA}{\kind}}{\expr}{\type}{\effpure} }
      { \typing{\env}{\lamD{\tvarA}{\kind}\expr}
          {\tforall{\tvarA}{\kind} \type}{\effpure} }

    \inferrule*[right=T-AppD]
      { \typing{\env}{\expr}{\tforall{\tvarA}{\kind} \type}{\effect}
        \and
        \kinding{\env}{\descriptor}{\kind} }
      { \typing{\env}{\expr\appD{\descriptor}}
          {\Subst{\tvarA}{\descriptor}{\type}}{\effect} }

    \inferrule*[right=T-Conv]
      { \typing{\env}{\expr}{\type_1}{\effect_1}
        \and
        \typeEquiv{\type_1}{\type_2}
        \and
        \effEquiv{\effect_1}{\effect_2} }
      { \typing{\env}{\expr}{\type_2}{\effect_2} }
  \end{mathpar}
  \caption{Typing rules of the declarative type system.}
  \label{fig:typing}
\end{figure}

Finally, we present the typing rules of the declarative type system
in \autoref{fig:typing}.
Most of the rules are standard, so we will only highlight a few details.
In the rule \textsc{T-LamU} for unannotated lambda abstractions,
we require that the argument type is monomorphic,
as the type inference algorithm is not able to infer polymorphic types.
In the rule \textsc{T-AppD} for type applications,
we check that the descriptor has the expected kind,
and use capture-avoiding substitution to substitute it
in the body of the polymorphic type (we use the notation $\bind{\usubst}{A}$
for applying a substitution $\usubst$ in the term~$A$).
The \textsc{T-Conv} rule allows to change the type and effect of an expression
to equivalent ones.

It is worth noting that premises requiring well-formedness
of the argument type in the rules \textsc{T-Lam} and \textsc{T-LamU}
(marked with \omitted{blue}) were omitted in the original presentation.
As such premises are standard, we assume that this was a minor oversight,
and include them in our presentation.

\section{Unification}

To support type inference, Jouvelot and Gifford
designed an unification algorithm that unifies types modulo
equivalence relation presented in the previous section.
The novelty of their approach is that unification produces a substitution
that equates the type structures,
but delays unification of effects by producing
a set of equality constraints that need to be solved later.

\begin{figure}[t]
  \begin{minipage}[t]{0.45\textwidth}
  \begin{align*}
  &\unify{\unifVarX}{\type} = \\
    &\algI \omitted{\algIfThen{\type = \unifVarX}
      \unifyRes{\varnothing}{\varnothing}} \\
    &\algI \omitted{\algIfThen{\unifVarX \in \FUV{\type}} \algFail} \\
    &\algI \algIfThen{\monotype{\type}}
      \unifyRes{\mkusubst{\unifVarX}{\type}}{\varnothing} \\
    &\algI \algElse \algFail \\
  &\unify{\type}{\unifVarX} =
    \unify{\unifVarX}{\type}
  \end{align*}
  \end{minipage}
  \begin{minipage}[t]{0.45\textwidth}
  \begin{align*}
  &\unify{\type_1 \tarrow{\effect_1} \type_1'}
         {\type_2 \tarrow{\effect_2} \type_2'} = \\
    &\algI \algLet{\unifyRes{\usubst_1}{\constrs_1}}
      {\unify{\type_1}{\type_2}} \\
    &\algI \algLet{\unifyRes{\usubst_2}{\constrs_2}}
      {\unify{\bind{\usubst_1}{\type_1'}}{\bind{\usubst_1}{\type_2'}}} \\
    &\algI \unifyRes
      {\usubst_2 \usubst_1}
      {\constrs_1 \cup \constrs_2 \cup \{ \mkconstr{\effect_1}{\effect_2} \}} \\
  &\unify{\tforall{\tvarA_1}{\kind} \type_1}
         {\tforall{\tvarA_2}{\kind} \type_2} = \\
    &\algI \algFresh{\tvarB} \\
    &\algI \unify
        {\USubst{\tvarA_1}{\tvarB}{\type_1}}
        {\USubst{\tvarA_2}{\tvarB}{\type_2}}
  \end{align*}
  \end{minipage}
  \caption{Algebraic unification algorithm.}
  \label{fig:unification}
\end{figure}

We present their unification algorithm in \autoref{fig:unification}.
The algorithm takes two types and returns a pair consisting of
a substitution for unification variables
and a set of constraints between effects.
The algorithm is said to be based on Robinson's
unification algorithm~\cite{DBLP:journals/jacm/Robinson65},
but the ``same variable case'' and occurs check
(both marked with \omitted{blue})
were omitted in the original presentation.
We assume that this was a minor oversight,
and include them in our presentation.

Since unification variables represent some unknown monomorphic types,
in the first, variable-and-type case, the algorithm additionally
checks if the type is monomorphic before producing the substitution.
To unify two arrows, the algorithm recursively unifies
types of their arguments and results, but delays unification of their effects
by adding a constraint to the result.
The last case handles unification of polymorphic types.
Since the two types may use different bound type variables,
the algorithm first renames both bound variables
to the same fresh variable $\tvarB$, and then recursively unifies
the bodies of the polymorphic types.
If none of the cases match, the algorithm fails.
Note that we use a different notation for substitutions in the algorithm
than in the declarative type system.
The difference will be explained in the next section.

\subsection{Semantics of Unification Variables}

To understand the unification algorithm,
we have to clarify the intended semantics of unification variables
that appear under type quantifiers, for example in types like
$\type \eqdef \tforall{\tvarA}{\ktype} \unifVarX \tarrow{\effpure} \tvarA$.
There are two possible semantics.
\begin{description}
\item[Capturing semantics.]
  Under capturing semantics,
  the unification variable can be instantiated to a type
  that contains bound type variables, resulting in variable capture.
  In the above example, the unification variable~$\unifVarX$
  can be instantiated to $\tvarA$, resulting in the type $\type$ being
  instantiated to $\tforall{\tvarA}{\ktype} \tvarA \tarrow{\effpure} \tvarA$.
  Proper handling of capturing semantics of unification is usually
  problematic, but possible thanks to explicit substitutions~\cite{%
    DBLP:journals/iandc/DowekHK00,%
    DBLP:conf/tphol/Huet02}
  or permutations of names~\cite{DBLP:journals/tcs/UrbanPG04}.
\item[Capture-avoiding semantics.]
  In capture-avoiding semantics, type variables cannot be captured
  by instantiating unification variables.
  The capture-avoiding semantics is much easier to implement
  when using standard capture-avoiding substitution.
  On the other hand, it is less expressive,
  so it endangers completeness of the type inference algorithm.
\end{description}

The original presentation does not state explicitly which semantics
is used. However, there are several clues.
First, the type inference algorithm is claimed to be complete,
so the more expressive capturing semantics is likely intended.
Second, as we will see in \autoref{sec:inference},
the type inference algorithm for the type application case
substitutes for a type variable that is not bound,
but (probably intentionally) leaks out of its scope.
This suggests that the actual identifiers used for bound type variables
matter, which is not consistent with capture-avoiding substitutions,
as capture-avoiding substitutions may change the names of bound variables.
Third, the presented part of the proof of completeness would be unsound
if capture-avoiding substitutions are assumed.

One of the authors of the original paper
confirmed in private communication
that ``he doesn't remember that `local' unification\footnote{
  \emph{i.e.}, unification with capturing semantics} was prohibited~\cite{%
  Jouvelot-communication}.''
Moreover, he stated that they intended to resolve the issue of variable capture
by assuming that all bound variables have distinct names in the whole program.
The latter is supported by the paper itself,
where it is assumed that the whole program
is ``alpha-renamed so there are no identifier name conflicts.''

To summarize, we conclude the following:
\begin{itemize}
\item the intended semantics of the unification variables is
  the capturing semantics,
\item scoping issues are resolved by alpha-renaming the whole program
  to ensure that all bound variables have distinct names,
  so it doesn't matter if we use capture-avoiding or capturing substitutions,
\item the type inference algorithm requires
  the names of bound variables to be preserved,
  thus it is more precise to use capturing substitutions,
\item the completeness proof appears to rely on capturing substitutions.
\end{itemize}
In order to keep our presentation precise,
and differentiate between capture-avoiding substitutions
(used by the declarative type system)
and capturing substitutions (used by the algorithm),
we use curly braces for capture-avoiding substitutions ($\mksubst{\tvarA}{\type}$)
and square brackets for capturing substitutions ($\mkusubst{\tvarA}{\type}$).

\subsection{Capturing Substitutions}

Some readers may feel uncomfortable with capturing substitutions,
as they are not commonly used
(at least in the metatheory of programming languages),
and may lead to unintended variable capture if not handled carefully.
To convince such readers that capturing substitutions
are not as bad as they may seem,
let us consider a simplified declarative type system
which forbids unannotated lambda abstractions.
For the purposes of this subsection we take the type system
from \autoref{fig:typing}, but remove the rule \textsc{T-LamU}.

For such a simplified type system,
it is possible to define a straightforward type inference algorithm
that doesn't use unification.
Let us consider a variant of such an algorithm
that uses capturing substitutions.
The two most interesting cases in the algorithm are the following.
\begin{center}
  \begin{minipage}[t]{0.45\textwidth}
    \begin{align*}
      & \sinfer{\env}{\expr_1\;\expr_2} = \\
      & \algI \algLet{\sinferRes{\type_a \tarrow{\effect} \type_1}{\effect_1}}
          {\sinfer{\env}{\expr_1}} \\
      & \algI \algLet{\sinferRes{\type_2}{\effect_2}}
          {\sinfer{\env}{\expr_2}} \\
      & \algI \algIfThen{\type_a \equiv \type_2}
          \sinferRes{\type_1}{\effect_1 \effjoin \effect_2 \effjoin \effect} \\
      & \algI \algElse \algFail
    \end{align*}
  \end{minipage}
  \begin{minipage}[t]{0.45\textwidth}
    \begin{align*}
      & \sinfer{\env}{\expr\appD{\descriptor}} = \\
      & \algI \algLet{\sinferRes{\tforall{\tvarA}{\kind} \type}{\effect}}
          {\sinfer{\env}{\expr}} \\
      & \algI \algIfThen{\kinding{\env}{\descriptor}{\kind}}
          \sinferRes{\USubst{\tvarA}{\descriptor}{\type}}{\effect} \\
      & \algI \algElse \algFail
    \end{align*}
  \end{minipage}
\end{center}
In the application case, we infer types of both the function and its argument,
make sure that the function has an arrow type, and check that
the argument type matches the expected type.
Since we do not have unification variables,
it is sufficient to check $\alpha$-equivalence of types
(modulo equational theory of effects).
In the type application case, we infer the type of the expression,
make sure it is a polymorphic type
that binds a type variable of the expected kind,
and then substitute the descriptor for the bound type variable,
using a capturing substitution.

The proof that this algorithm is sound and complete
with respect to the simplified declarative type system
is quite technical, so we sketch only the main ideas here.
First, we define the following two notions of uniqueness of bound variables.
\begin{definition}
  For a given term $A$ of any syntactic category
  (\emph{e.g.}, expression, type, typing environment),
  we say that it has \emph{globally unique bound variables} with respect to
  the environment $\env$ (written $\gUnique{\env}{A}$)
  if for each bound type or effect variable $\tvarA$ in $A$,
  it is bound only once in $A$,
  and is not in the domain of the environment $\env$
  (\emph{i.e.}, $(\kindExt{\tvarA}{\kind}) \notin \env$ for any kind $\kind$).
  We say that $A$ has \emph{locally unique bound variables} with respect to
  the environment $\env$ (written $\lUnique{\env}{A}$)
  if each bound type or effect variable $\tvarA$ in $A$
  does not shadow any other bound variable in $A$,
  and is not in the domain of the environment $\env$.
  In particular, a typing environment $\env$ has locally unique bound variables
  with respect to itself ($\lUnique{\env}{\env}$) if for each type assignment
  $(\varX : \type) \in \env$, we have $\lUnique{\env}{\type}$.
\end{definition}

\begin{example}
  Consider the environment $\env =
    [\kindExt{\tvarA}{\ktype},
     \varX : \tforall{\tvarB}{\ktype} \tvarB]$.
  \begin{itemize}
  \item We have $\gUnique{\varnothing}
      {\tforall{\tvarA}{\ktype} \tvarA \tarrow{\effpure} \tvarA}$,
    and $\lUnique{\varnothing}
      {\tforall{\tvarA}{\ktype} \tvarA \tarrow{\effpure} \tvarA}$,
    but none of $\gUnique{\env}
      {\tforall{\tvarA}{\ktype} \tvarA \tarrow{\effpure} \tvarA}$
    or $\lUnique{\env}
      {\tforall{\tvarA}{\ktype} \tvarA \tarrow{\effpure} \tvarA}$
    hold, because $\tvarA$ is in the domain of $\env$.
  \item We have $\gUnique{\env}{\tforall \tvarB{\ktype} \tvarB}$,
    and $\lUnique{\env}{\tforall \tvarB{\ktype} \tvarB}$,
    even though $\tvarB$ is bound in both $\env$ and the type.
  \item We have $\lUnique{\env}
      {(\tforall{\tvarB}{\ktype} \tvarB) \tarrow{\effpure}
       (\tforall{\tvarB}{\ktype} \tvarB)}$,
    because the two bindings of $\tvarB$
    do not shadow each other.
    On the other hand, $\gUnique{\env}
      {(\tforall{\tvarB}{\ktype} \tvarB) \tarrow{\effpure}
       (\tforall{\tvarB}{\ktype} \tvarB)}$
    does not hold,
    because $\tvarB$ is bound twice.
  \item None of $\gUnique{\env}
      {\tforall{\tvarC}{\ktype} \tvarC \tarrow{\effpure}
        \tforall{\tvarC}{\ktype} \tvarC}$
    or $\lUnique{\env}
      {\tforall{\tvarC}{\ktype} \tvarC \tarrow{\effpure}
        \tforall{\tvarC}{\ktype} \tvarC}$
    hold, because $\tvarC$ is bound twice, and the second binding
    shadows the first one.
  \item We have $\lUnique{\env}{\env}$,
    because we have $\lUnique{\env}{\tforall{\tvarB}{\ktype} \tvarB}$.
  \end{itemize}
\end{example}

With these notions, we can state the correctness lemma that says that
the capturing substitution coincides with the capture-avoiding substitution
when certain uniqueness conditions are met.
Since types returned by the type inference algorithm
don't necessarily have globally unique bound variables,
the lemma is stated for types with locally unique bound variables.

\begin{lemma}
  \label{lem:subst-equiv}
  If $\lUnique{\env, \kindExt{\tvarA}{\kind}}{\type}$
  and $\kinding{\env}{\descriptor}{\kind}$ hold,
  then $\Subst{\tvarA}{\descriptor}{\type} \equiv 
       \USubst{\tvarA}{\descriptor}{\type}$.
  Moreover, if we additionally have $\lUnique{\env}{\descriptor}$,
  then $\lUnique{\env}{\Subst{\tvarA}{\descriptor}{\type}}$ holds.
\end{lemma}

To make sure that requirements of the lemma are satisfied when
substitution is used in the type inference algorithm,
we can prove the following invariant of the algorithm.

\begin{lemma}
  Assume that $\sinfer{\env}{\expr} = \sinferRes{\type}{\effect}$
  and the input of the type inference algorithm
  satisfies the following conditions:
  \begin{itemize}
  \item $\gUnique{\env}{\expr}$,
  \item $\lUnique{\env}{\env}$,
  \item $\env$ and $\expr$ have disjoint sets of bound variables.
  \end{itemize}
  Then the following conditions hold:
  \begin{itemize}
  \item $\lUnique{\env}{\type}$,
  \item each bound variable in $\type$ is bound in $\expr$ or in $\env$.
  \end{itemize}
\end{lemma}

With this invariant and \autoref{lem:subst-equiv},
the proofs of soundness and completeness
of the type inference algorithm become straightforward.

\subsection{Criticism}
\label{sec:unif-criticism}

In the original presentation,
the unification algorithm is said to be correct.
Since the algorithm does not unify effects directly, but rather
produces constraints between them,
the correctness statement uses the following notion of a model.
\begin{definition}[Model]
  We say that a capturing substitution $\umodel$
  is a \emph{model}
  if it maps effect unification variables to \emph{ground} effects.
  We say that a model $\umodel$ models a constraint set $\constrs$
  (written $\umodel \models \constrs$)
  if for each constraint $\mkconstr{\effect_1}{\effect_2} \in \constrs$,
  we have $\effEquiv{\bind{\umodel}{\effect_1}}{\bind{\umodel}{\effect_2}}$.
\end{definition}
It is not clear to us what is meant by \emph{ground} effects here.
Certainly, ground effects cannot contain unification variables,
but can they contain regular effect variables or constants?
However, this matter is not crucial for the problems we identified.

The original correctness theorem is stated as follows.
\begin{theorem}[Correctness, Theorem~1 in \cite{DBLP:conf/popl/JouvelotG91}]
  If $\unify{\type_1}{\type_2} = \unifyRes{\usubst}{\constrs}$
  and $\umodel \models \constrs$ for some model $\umodel$,
  then $\typeEquiv
    {\bind{\umodel}{\bind{\usubst}{\type_1}}}
    {\bind{\umodel}{\bind{\usubst}{\type_2}}}$.
\end{theorem}

There are no details of the proof, but it is said
that the proof follows by induction on the structure of types.
We have found that the unification algorithm behaves incorrectly
in the polymorphic type case.
Consider the following example.
\begin{eqnarray*}
  \unify{\tforall{\tvarA}{\ktype} \tvarA}
        {\tforall{\tvarA}{\ktype} \unifVarX} & = &
     \unify{\USubst{\tvarA}{\tvarB}{\tvarA}}
          {\USubst{\tvarA}{\tvarB}{\unifVarX}} \\
  & = & \unify{\tvarB}{\unifVarX} \\
  & = & \unifyRes{\mkusubst{\unifVarX}{\tvarB}}{\varnothing}
\end{eqnarray*}
The algorithm returns an empty set of constraints,
which is satisfied by any model, in particular by
the identity substitution.
However, if we apply the substitution to the input types, we get
types $\tforall{\tvarA}{\ktype} \tvarA$ and $\tforall{\tvarA}{\ktype} \tvarB$,
which are not equivalent.
A similar counterexample can be constructed for effect-polymorphic types:
$\tforall{\tvarA}{\keffect} \type_1 \tarrow{\tvarA} \type_2$
and
$\tforall{\tvarA}{\keffect} \type_1 \tarrow{\unifVarX} \type_2$.
Moreover, the problem remains even if we assume capture-avoiding
semantics of unification variables.

Incorrectness of the unification algorithm does not directly imply
incorrectness of the type inference algorithm,
since the user cannot write types with unification variables directly.
However, as we will see in the next section,
such types can be constructed internally by the type inference algorithm.

\section{Inference Algorithm}
\label{sec:inference}

\begin{figure}[t]
  \begin{minipage}[t]{0.48\textwidth}
  \begin{align*}
    & \Infer{\env}{\varX} =
      \InferRes{\env(\varX)}{\effpure}{\varnothing}{\varnothing} \\
    & \Infer{\env}{\lam{\varX}{\type}\expr} = \\
    & \algI \omitted{\checkWF{\env}{\type}{\ktype};} \\
    & \algI \algLet{\InferRes{\type'}{\effect}{\usubst}{\constrs}}
        {\Infer{\env, \varX : \type}{\expr}} \\
    & \algI \InferRes
        {\type \tarrow{\effect} \type'}
        {\effpure}
        {\usubst}
        {\constrs} \\
    & \Infer{\env}{\lamU{\varX}\expr} = \\
    & \algI \algFresh{\unifVarX} \\
    & \algI \algLet{\InferRes{\type}{\effect}{\usubst}{\constrs}}
        {\Infer{\env, \varX : \unifVarX}{\expr}} \\
    & \algI \InferRes
        {\usubst(\unifVarX) \tarrow{\effect} \type}
        {\effpure}
        {\usubst}
        {\constrs} \\
    & \Infer{\env}{\expr_1\;\expr_2} = \\
    & \algI \algLet{\InferRes{\type_1}{\effect_1}{\usubst_1}{\constrs_1}}
        {\Infer{\env}{\expr_1}} \\
    & \algI \algLet{\InferRes{\type_2}{\effect_2}{\usubst_2}{\constrs_2}}
        {\Infer{\bind{\usubst_1}{\env}}{\expr_2}} \\
    & \algI \algFresh{\unifVarX_t, \unifVarX_e} \\
    & \algI \algLet{\unifyRes{\usubst_3}{\constrs_3}}
        {\unify
          {\bind{\usubst_2}{\type_1}}
          {\type_2 \tarrow{\unifVarX_e} \unifVarX_t}} \\
    & \algI \InferRes
        {\usubst_3(\unifVarX_t)}
        {\effect_1 \effjoin \effect_2 \effjoin \unifVarX_e}
        {\usubst_3 \usubst_2 \usubst_1}
        {\constrs_1 \cup \constrs_2 \cup \constrs_3}
  \end{align*}
  \end{minipage}
  \quad
  \begin{minipage}[t]{0.48\textwidth}
  \begin{align*}
    & \Infer{\env}{\lamD{\tvarA}{\kind}\expr} = \\
    & \algI \algLet{\InferRes{\type}{\effect}{\usubst}{\constrs}}
        {\Infer{\env, \kindExt{\tvarA}{\kind}}{\expr}} \\
    & \algI \algLet{\constrs'}
        {\constrs \cup \{ \mkconstr{\effect}{\effpure} \}} \\
    & \algI \algLet{\{\ident_i\}_{i=1\ldots n}}
        {\FV{\constrs'} \setminus
          \FV{\bind{\usubst}{\env, \kindExt{\tvarA}{\kind}}}} \\
    & \algI \algFresh{\tvarB} \\
    & \algI \algFresh{\unifVarX_1, \ldots, \unifVarX_n} \\
    & \algI \InferRes
        {\tforall{\tvarA}{\kind}\type}
        {\effpure}
        {\usubst}
        {\bind
          {(\mkusubst{\tvarA}{\tvarB}
            \overline{\mkusubst{\ident_i}{\unifVarX_i}})}
          {\constrs'} \cup
         \constrs'} \\
    & \Infer{\env}{\expr\appD{\descriptor}} = \\
    & \algI \algLet{\InferRes{\type'}{\effect}{\usubst}{\constrs}}
        {\Infer{\env}{\expr}} \\
    & \algI \algMatch{\type'} \\
    & \algI \algCase{\tforall{\tvarA}{\kind} \type} \\
    & \algI \algI \omitted{\checkWF{\env}{\descriptor}{\kind};} \\
    & \algI \algI \InferRes
        {\USubst{\tvarA}{\descriptor}{\type}}
        {\effect}
        {\usubst}
        {\USubst{\tvarA}{\descriptor}{\constrs}} \\
    & \algI \algCase{\_} \algFail
  \end{align*}
  \end{minipage}
  \caption{Type inference algorithm.}
  \label{fig:infer}
\end{figure}

The type inference algorithm is presented in \autoref{fig:infer}.
It is assumed that the input expression does not contain unification variables,
and all bound type variables have distinct names.
The algorithm returns a tuple consisting of the inferred type, effect,
a substitution for type unification variables,
and a set of constraints between effects.\footnote{%
  We changed the order of the last two components of the result
  to be consistent with the unification algorithm.}

The cases for variables, both kinds of lambda abstractions,
and function application are straightforward.
We only note that in the annotated lambda case,
the check for well-formedness of the type
was omitted in the original presentation.
This omission was likely intentional,
as the algorithm assumes that all bound type variables are distinct,
so this well-formedness check might be done
during earlier phase of the alpha-renaming of the program.

In the type abstraction case,
after inferring the type and effect of the body,
the algorithm first adds a constraint $\mkconstr{\effect}{\effpure}$
to the set of constraints, to enforce purity restriction.
Then, the algorithm ensures that the obtained set of constraints
is polymorphic with respect to the variable~$\tvarA$.
To do so, it simulates instantiation of the polymorphic function,
by including copy of all constraints with~$\tvarA$ substituted
by a fresh constant~$\tvarB$.
Additionally, to make sure that the check for polymorphism
would not affect propagated constraints~$\constrs'$,
all variables that do not appear in the environment
(in the domain or in types assigned to variables)
are replaced with fresh unification variables.
Note that the bound variable~$\tvarA$ may escape its scope through
constraints~$\constrs'$, and this is likely intentional,
as we will see in the type application case.

The case for type application is straightforward, except two things.
First, the well-formedness check for the descriptor
was omitted in the original presentation.
Similarly to the annotated lambda case, this omission was likely intentional.
Second, the algorithm substitutes the descriptor~$\descriptor$
for the bound type variable~$\tvarA$ not only in the body
of the polymorphic type, but also in the set of constraints.
The purpose of this substitution is not explained in the original presentation
and is not clear to us.
However, this substitution is likely intentional,
as it is repeated in the proof of soundness.
From the proof we were unable to deduce the reason for this substitution,
but it seems that it is related to the fact
that bound type variables escape their scope through constraints
in the type abstraction case.
However, the presented (part of the) proof would become simpler
if this substitution was removed.
Moreover, the variable~$\tvarA$ in this strange substitution
comes from opening the binder~$\tforall{\tvarA}{\kind} \type$,
which suggests that actual names of bound variables matter
and the capturing semantics of unification variables is intended.

\subsection{Soundness}

The presented type inference algorithm is said to be sound
with respect to the declarative type system.
The soundness theorem is stated as follows.
\begin{theorem}[Soundness, Theorem~3 in \cite{DBLP:conf/popl/JouvelotG91}]
  If $\Infer{\env}{\expr} =
    \InferRes{\type}{\effect}{\usubst}{\constrs}$
  and $\umodel \models \constrs$ for some model $\umodel$,
  then
  $\typing
    {\bind{\umodel}{\bind{\usubst}{\env}}}
    {\expr}
    {\bind{\umodel}{\type}}
    {\bind{\umodel}{\effect}}$.
\end{theorem}
The presented proof focuses only on the application and type application cases.
However, we have found that other cases, not covered by the proof,
are problematic.
Below, we summarize the issues we have identified.
One of them leads to a counterexample that shows
that the presented theorem does not hold.

\paragraph{Incorrect unification.}

As we have seen in \autoref{sec:unif-criticism},
the unification algorithm is incorrect in handling polymorphic types.
It turns out that instances similar to the counterexample
presented in that section may appear in the type inference algorithm.
Consider the following program.
\[
  \lam{\varF}{
    (\tforall{\tvarA}{\ktype} \tvarA \tarrow{\effpure} \tvarA)
    \tarrow{\effpure} \type}
  \varF\;
  (\lamD{\tvarB}{\ktype} \lamU{\varX} \varX)
\]
For this program, the type inference algorithm triggers a unification
of types $\tforall{\tvarA}{\ktype} \tvarA \tarrow{\effpure} \tvarA$
and $\tforall{\tvarB}{\ktype} \unifVarX \tarrow{\effpure} \unifVarX$,
which leads to the same incorrect behavior
as in \autoref{sec:unif-criticism}.
However, in this case, the whole program is typable
and the whole type inference algorithm returns the correct type.
Without extending the calculus we were unable to construct
a counterexample for the soundness theorem based on this issue alone.

\paragraph{Escaping variables in the type abstraction case.}

In the type abstraction case, the algorithm does not ensure
that the bound variable~$\tvarA$ does not escape its scope.
This makes the type inference algorithm unsound.
For example, consider the following program.
\[
  \lamU{\varF} \lamD{\tvarA}{\ktype} \lam{\varX}{\tvarA} \varF\;\varX
\]
This program is not typable in the declarative type system,
but the type inference algorithm accepts it, performing the following steps
(for the sake of readability, we simplify joins with the pure effect).
\begin{itemize}
\item First, it creates a fresh unification variable~$\unifVarZ$ that
  represents the type of~$\varF$.
\item Then, for the innermost lambda abstraction,
  the algorithm infers the type $\tvarA \tarrow{\unifVarX} \unifVarY$
  and the substitution
  $\mkusubst{\unifVarZ}{\tvarA \tarrow{\unifVarX} \unifVarY}$,
  where $\unifVarX$ and $\unifVarY$ are fresh unification variables
  created in the application case.
\item For the type abstraction,
  the algorithm returns the type
  $\tforall{\tvarA}{\ktype} \tvarA \tarrow{\unifVarX} \unifVarY$,
  the same substitution, and two copies of a trivial constraint
  $\mkconstr{\effpure}{\effpure}$.
  Note that the bound variable~$\tvarA$
  escapes its scope through the substitution.
\item Finally, the algorithm returns the type
  $(\tvarA \tarrow{\unifVarX} \unifVarY)
    \tarrow{\effpure}
    \tforall{\tvarA}{\ktype} \tvarA \tarrow{\unifVarX} \unifVarY$,
  and the same set of constraints that are trivially satisfied.
\end{itemize}

It is possible to construct a similar counterexample, where
effect variables escape their scope:
\[
  \lamU{\varF} \lamD{\tvarA}{\keffect}
    \lam{\varX}{\type \tarrow{\tvarA} \type} \varF\;\varX
  \textrm{,}
\]
where $\type$ is some base type.
While fixing the algorithm to prevent escaping of type variables
is rather easy,
escaping effect variables seems to be inherent to the design of the algorithm:
the algorithm only collects constraints between effects,
and does not analyze them until the very end of type inference.

\subsection{Completeness}

The type inference algorithm is also said to be complete
with respect to the declarative type system.
The completeness theorem is stated as follows.
\begin{theorem}[Completeness, Theorem~4 in \cite{DBLP:conf/popl/JouvelotG91}]
  If $\typing{\bind{\umodel}{\bind{\usubst}{\env}}}{\expr}{\type}{\effect}$
  then $\Infer{\env}{\expr} =
    \InferRes{\type'}{\effect'}{\usubst'}{\constrs'}$
  for some $\type'$, $\effect'$, $\usubst'$, and $\constrs'$
  such that there exists a model $\umodel'$ and a substitution $\usubst''$
  satisfying the following conditions:
  \begin{itemize}
  \item $\bind{\umodel'}{\bind{\usubst''}{\bind{\usubst'}{\env(\varX)}}}
      =
      \bind{\umodel}{\bind{\usubst}{\env(\varX)}}$
      for each $\varX \in \FV{\expr}$,
  \item $\umodel' \models \constrs'$,
  \item $\typeEquiv{\type}{\bind{\umodel'}{\bind{\usubst''}{\type'}}}$,
  \item $\effEquiv{\effect}{\bind{\umodel'}{\effect'}}$.
  \end{itemize}
\end{theorem}
The presented proof focuses only on the lambda abstraction
and polymorphic abstraction cases.
As the case for polymorphic abstraction ($\lamD{\tvarA}{\kind}\expr$)
concerns variable binding and substitutions,
let us look at the following part of the proof in more detail
(we changed the notation to be consistent with our presentation).

``[...] By induction,
there exist $\InferRes{\type_0}{\effect_0}{\usubst_0}{\constrs_0}$,
$\umodel_0$ and $\usubst_0'$ that verify the theorem for~$\expr$
in the environment~$\bind{\umodel}{\bind{\usubst}
  {\env, \kindExt{\tvarA}{\kind}}}$.
[...]
To prove the theorem, pick:
\begin{align*}
  \type'    & = \tforall{\tvarA}{\kind} \type_0 &
  \effect'  & = \effpure &
  \usubst'  & = \usubst_0 \\
  \constrs' & = \textrm{[...]} &
  \umodel'  & = \umodel_0[\overline{\unifVarX_i \mapsto \umodel_1(\ident_i)}] &
  \usubst'' & = \usubst_0'
\end{align*}
[...]. The theorem is trivially verified, except for the verification of
constraints. [...]''

Let us verify the third condition of the theorem (for types)
using the above choices.
We have to show that
\[
  \typeEquiv
    {\tforall{\tvarA}{\kind} \type}
    {\bind{\umodel'}{\bind{\usubst''}
      {\tforall{\tvarA}{\kind} \type_0}}}
  \textrm{.}
\]
By induction hypothesis, we have
$\typeEquiv{\type}{\bind{\umodel_0}{\bind{\usubst_0}{\type_0}}}$,
and we can ignore $[\overline{\unifVarX_i \mapsto \umodel_1(\ident_i)}]$
of $\umodel'$, as unification variables $\unifVarX_i$
do not appear in $\type_0$.
Thus, we have to show that
\[
  \typeEquiv
    {\tforall{\tvarA}{\kind} \bind{\umodel_0}{\bind{\usubst_0}{\type_0}}}
    {\bind{\umodel_0}{\bind{\usubst_0} {\tforall{\tvarA}{\kind} \type_0}}}
  \textrm{.}
\]
Since we use capturing substitutions, this equivalence trivially holds.
However, if the capture-avoiding substitutions were intended,
then we would have to show that variable~$\tvarA$ does not appear
in the substitution~$\usubst_0$.
There is no guarantee of this, so the proof would be incorrect in that case.
Thus, we have another indication that capturing semantics
of unification variables is intended.

However, we have found that the completeness theorem does not hold.
The problems lie in the function application
and polymorphic application cases, both omitted in the presented proof.
While the first issue is rather easy to fix,
the second one is more serious.
We discuss both issues below.

\paragraph{Arrow types with polymorphic results.}

In the function application case,
the algorithm unifies the type of the function
with the type $\type_2 \tarrow{\unifVarX_e} \unifVarX_t$,
where $\type_2$ is the type of the argument,
and $\unifVarX_e$ and $\unifVarX_t$ are fresh unification variables.
But type unification variables can represent only monomorphic types,
so the algorithm disallows applying functions with polymorphic result types.
An analogous restriction is not present in the declarative type system,
so the type inference algorithm is incomplete.
As a counterexample, consider the following program.
\[
  (\lamU{\varX}\lamD{\tvarA}{\ktype} \varX)\;42
\]
This program is typable in the declarative type system,
but the algorithm would fail on unifying
$\unifVarY \tarrow{\effpure} \tforall{\tvarA}{\ktype} \unifVarY$
with $\tInt \tarrow{\unifVarX_e} \unifVarX_t$,
which is required in the application case.

This issue has a straightforward fix.
Similarly to the type application case,
we could pattern-match the type of the function and
unify only the types of arguments.

\paragraph{Substitution before instantiation.}

The presented algorithm assumes capturing semantics of unification variables.
This means that unification variables that appear under type quantifier
can be instantiated to types containing a variable bound by that quantifier.
However, in the type application case,
when instantiating a polymorphic type $\tforall{\tvarA}{\kind} \type$,
the algorithm substitutes a descriptor for the bound variable~$\tvarA$
in the type~$\type$,
ignoring the fact that unification variables in~$\type$
may represent types containing~$\tvarA$.

Consider the unannotated polymorphic identity function
$\lamD{\tvarA}{\ktype} \lamU{\varX} \varX$.
For this function,
the algorithm infers the type
$\tforall{\tvarA}{\ktype} \unifVarX \tarrow{\effpure} \unifVarX$,
where $\unifVarX$ is a fresh unification variable.
Notice that the body of the polymorphic type does not contain~$\tvarA$
directly, but $\unifVarX$ may be instantiated to a type containing~$\tvarA$.
Now, consider applying this function to some type~$\type$.
The algorithm will substitute~$\type$ for~$\tvarA$, but this substitution
will do nothing! The type of such an application will be inferred
as~$\unifVarX \tarrow{\effpure} \unifVarX$.

Constructing a counterexample based on this issue is not
straightforward, because in the presented calculus,
each polymorphic function that has unification variables in its type
can be used at most once (applied,
or passed as an argument to another function).
However, observe that $\unifVarX$ stands for a monomorphic type,
but polymorphic types can be instantiated to any type descriptor,
even polymorphic ones. Consider the following program.
\[
  (\lamD{\tvarA}{\ktype} \lamU{\varX} \varX)
  \appD{\tforall{\tvarB}{\ktype} \tvarB \tarrow{\effpure} \tvarB}\;
  (\lamD{\tvarC}{\ktype} \lamU{\varY} \varY)
\]
This program is typable in the declarative type system,
but the type inference algorithm fails, trying to instantiate
unification variable with a polymorphic type.

A more straightforward counterexample could be constructed
if we extended the calculus with
let-bindings~($\elet{\varX}{\expr_1}{\expr_2}$),
with the usual typing rule, and the following extension
to the type inference algorithm.\footnote{%
  At the end of the original paper, the authors consider
  a more general form of let-bindings, that allows for
  some form of let-polymorphism. However, for our purposes,
  it is sufficient to consider only its simpler form, assuming
  that all expressions are \emph{expansive}.}
\begin{align*}
  & \Infer{\env}{\elet{\varX}{\expr_1}{\expr_2}} = \\
  & \algI \algLet{\InferRes{\type_1}{\effect_1}{\usubst_1}{\constrs_1}}
      {\Infer{\env}{\expr_1}} \\
  & \algI \algLet{\InferRes{\type_2}{\effect_2}{\usubst_2}{\constrs_2}}
      {\Infer{\bind{\usubst_1}{\env}, \varX : \type_1}{\expr_2}} \\
  & \algI \InferRes
      {\type_2}
      {\effect_1 \effjoin \effect_2}
      {\usubst_2 \usubst_1}
      {\constrs_1 \cup \constrs_2}
\end{align*}

Now, consider the following program.
\begin{align*}
  & \lamD{\tvarA_1}{\ktype} \lamD{\tvarA_2}{\ktype}
    \lam{\varX_1}{\tvarA_1} \lam{\varX_2}{\tvarA_2} \\
  & \quad \elet{\varF}{\lamD{\tvarB}{\ktype} \lamU{\varY} \varY} \\
  & \quad \elet{\varZ}{\varF \appD{\tvarA_1} \; \varX_1} \\
  & \quad \varF \appD{\tvarA_2} \; \varX_2
\end{align*}
Again, this program is typable in the declarative type system.
However, the type inference algorithm fails, performing the following steps.
\begin{itemize}
\item First, it assigns the type
  $\tforall{\tvarB}{\ktype} \unifVarX \tarrow{\effpure} \unifVarX$
  to~$\varF$, where $\unifVarX$ is a fresh unification variable.
\item The type application $\varF \appD{\tvarA_1}$
  results in the type $\unifVarX \tarrow{\effpure} \unifVarX$,
  as in the previous example.
\item The applying $\varF \appD{\tvarA_1}$ to~$\varX_1$
  leads to instantiation of the unification variable~$\unifVarX$
  with type~$\tvarA_1$.
  Now, the type of~$\varF$ is
  $\tforall{\tvarB}{\ktype} \tvarA_1 \tarrow{\effpure} \tvarA_1$.
  It does not depend on~$\tvarB$ anymore.
\item Finally, the application of $\varF \appD{\tvarA_2}$ to~$\varX_2$
  fails, because the type of~$\varX_2$ does not match~$\tvarA_1$.
\end{itemize}

A similar counterexample can be constructed based on effect polymorphism,
and produces unsatisfiable effect constraints instead of a type mismatch.

\section{Conclusion}

In the presence of higher-rank polymorphism, it is rather easy for a type
inference algorithm to look entirely correct while containing a number of
subtle variable-binding bugs.
These issues are also easy to overlook in a pen-and-paper proof, in which
manipulations on variables and substitutions are seen as more of a chore than
an important part of the reasoning.
Such problems can be prevented with the aid of proof assistants, but it still
took a long time to develop techniques for dealing with variable binding
precisely, yet without tedium.
Several formalizations of type inference for sophisticated calculi were
developed since~\cite{%
  DBLP:conf/itp/ZhaoOS18,%
  DBLP:journals/pacmpl/ZhaoOS19,%
  DBLP:conf/ecoop/ZhaoO22,%
  DBLP:journals/pacmpl/CuiJO23,%
  DBLP:conf/itp/BosmanKS23,%
  DBLP:journals/pacmpl/JiangCO25,%
  DBLP:journals/pacmpl/XueO24,%
  DBLP:journals/pacmpl/FanXX25,%
  BalikJP26}, which were feasible due to precise tracking of
variable scopes.
It is worth mentioning that this level of precision is not exclusive to
mechanized proofs, and could also be practiced in traditional pen-and-paper
proofs as well~\cite{kennedy1996type}.
It is our hope that with increasing adoption of proof assistants, such degree
of precision becomes the standard in this area of research.

\bibliographystyle{ACM-Reference-Format}
\bibliography{references}

\end{document}